\documentstyle[aaspp4]{article}
\begin{document}
\title{The Reappearance of the Transient Low Mass X-ray Binary X1658$-$298} 
\author{Stefanie Wachter}  
\affil{Cerro Tololo Inter-American Observatory, National Optical Astronomy 
Observatory\altaffilmark{1}, Casilla 603, La Serena, Chile}
\altaffiltext{1}{Operated by the Association of Universities for Research in 
Astronomy, Inc., under cooperative agreement with the National Science 
Foundation.}
\authoremail{swachter@noao.edu}
\author{Alan P. Smale}
\affil{Laboratory for High Energy Astrophysics/USRA, Code 662, NASA/GSFC, 
Greenbelt, MD 20771}
\authoremail{alan@osiris.gsfc.nasa.gov}
\author{Charles Bailyn}
\affil{Yale University, Dept. of Astronomy, PO Box 208101, New Haven CT 06520-8101}
\authoremail{bailyn@astro.yale.edu}

\begin{abstract}

In April 1999 the transient low mass X-ray binary X1658$-$298 resumed
its strong and persistent X-ray emission after a 21-year interval of
quiescence. We present RXTE data obtained soon after the reappearance,
including four eclipses with a mean duration of 901.9$\pm$0.8~sec and
ingress/egress times of 6--13~sec. Our updated ephemeris for the source
indicates that the 7.1-hr orbital period of the system is decreasing
with a timescale of 10$^7$ yr. Contemporaneous optical observations
provide the first-ever lightcurve of V2134 Oph, the 
optical counterpart of X1658$-$298.  
The optical modulation is highly variable from night to night and 
exhibits a distinct, narrow eclipse feature of about 0.2~mag superposed on
a gradual brightness variation with $\sim 0.7-0.8$~mag amplitude.
Our data indicate that
there is no significant offset between the time of mid-eclipse in the
X-ray and optical and that the
narrow optical eclipse feature is of the same duration as the
X-ray eclipse. This implies an accretion disk structure characterized
by enhanced optical emission coincident with the central X-ray emitting area.

\end{abstract}


\section{Introduction}

X1658$-$298 is a soft X-ray transient discovered in 
1976 by Lewin, Hoffmann, \& Doty (1976). The detection of type~I bursts 
indicates that the compact object in the system is a neutron star.
Observations during a temporary
brightening of the source in 1978 showed dips in the X-ray lightcurve. 
Detailed analysis of the combined 1976--1978 data set by Cominsky \&
Wood (1984, 1989) revealed that X1658$-$298 is one of the rare low mass
X-ray binary systems (LMXBs) that exhibits eclipses of the
central X-ray source by the mass donating star. 
The dipping activity lasts for about 25\% of the 7.1~hour
orbital cycle followed by an eclipse of $\sim 15$~min duration. 

The optical counterpart of X1658$-$298 was identified during the 1978 
X-ray outburst
with a faint ($V=18.3$), blue star (V2134~Oph) by Doxsey et al.\ (1979).
Spectroscopic observations show a typical LMXB spectrum, a blue continuum with
emission lines of \ion{He}{2} $\lambda$4686 and the \ion{C}{3}/\ion{N}{3}
$\lambda$4640/4650 blend (Canizares, McClintock, \& Grindlay 1979).
X1658$-$298
entered an X-ray off--state in 1979 and the counterpart 
became undetectable with a magnitude limit of $V>23$ (Cominsky, Ossmann, \&
Lewin 1983).

Renewed X-ray activity from X1658$-$298 was detected by
BeppoSAX on April 2--3 1999 (In't Zand et al.\ 1999), marking the first
X-ray detection of the source since 1978. Follow-up observations were
quickly scheduled with RXTE under a public Target of Opportunity
program, and with optical telescopes at CTIO.  In this paper we
present the first optical light curve of X1658$-$298, and the results
of our RXTE eclipse timing and spectral fitting analysis.

\section{Observations}

\subsection{X--ray}

X1658$-$298 was observed with the RXTE satellite for a series of four
public observations between 1999 April 5--15, soon after the
recommencement of X-ray activity. The X-ray data we present here were
obtained using the RXTE Proportional Counter Array (PCA) instrument
with the Standard 2 and E\_125us\_64M\_0\_1s configurations, with time
resolutions of 16~sec and 125~$\mu$sec, respectively.  The PCA consists
of five Xe proportional counter units (PCUs), with a combined
effective area of about 6500 cm$^2$ (Jahoda et al.\ 1996). For
operational reasons, differing numbers of PCUs were utilized in each
observation.  In Table~1 we list the observation times and the PCUs on
during each observation.  Data extraction was performed using the RXTE
standard analysis software, FTOOLS v4.2. The {\tt "skyvle/skyactiv"}
models generated by the RXTE PCA team were used for background
subtraction, and found to be accurate to better than 1 cs$^{-1}$.
Light curves and spectra were analyzed in the 2--20 keV band.
Barycentric corrections have been applied to all X-ray timings.  2\%
systematic errors were added to the spectral data before fitting, to
represent the current uncertainties in response matrix generation.

\begin{deluxetable}{lll}
\tablewidth{0pc}
\tablecaption{RXTE Observation Log}
\tablehead{
\colhead{Obs} &
\colhead{Start/Stop Time (UT)} &
\colhead{PCUs on}
}
\startdata
1 & 1999 Apr 5 20:12 -- Apr 6 01:18  & 0124 \nl
2 & 1999 Apr 9 19:34 -- Apr 9 20:13  & 023 \nl
3 & 1999 Apr 13 16:07 -- Apr 13 16:46 & 134 \nl
4 & 1999 Apr 15 17:58 -- Apr 15 18:26 & 13 \nl
\enddata
\end{deluxetable}

\subsection{Optical}

CCD $V$ and $I_C$ band photometry of V2134~Oph 
was performed with the CTIO  
1.5m and YALO telescope from UT 1999 April 29 to May 3. The image scale
at the telescopes was 0.24\arcsec\ pix$^{-1}$ and 0.30\arcsec~pix$^{-1}$,
respectively. 
The data were overscan corrected, bias corrected and flat-fielded in 
the standard 
manner using IRAF. Photometry was performed by point spread function fitting 
with DAOPHOT~II (Stetson 1993). The instrumental magnitudes were transformed to 
the standard system through comparison with previously calibrated local 
standards 
(Wachter \& Smale 1998). The intrinsic 1$\sigma$ error of the relative
photometry is about $\pm 0.02$~mag as derived from the rms scatter in the 
lightcurve of comparison stars of similar brightness.  The standardized 
magnitudes are 
accurate to about $\pm 0.10$~mag. Exposure times 
were 300~sec for the YALO data and 200 to 240~sec (around the times of eclipse)
for the 1.5m data,
depending on the observing conditions.

\section{Results}

\subsection{X-ray}

\begin{figure}[htbp]
\plotone{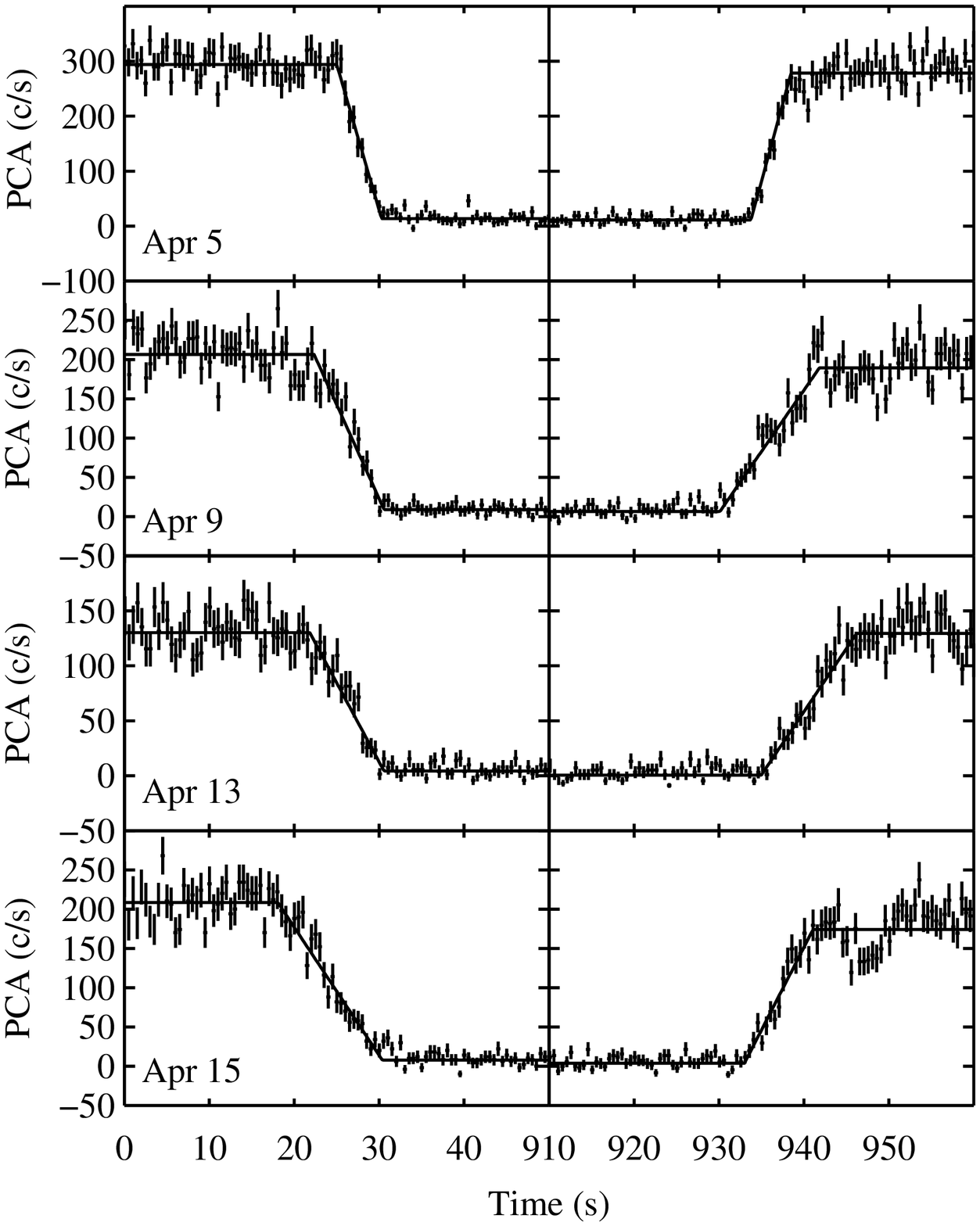}
\caption[]{The eclipse transitions observed from X1658$-$298 with the
RXTE PCA, in the 2--10 keV energy range, with 0.5~sec time resolution.
The data have been background-subtracted, and the best-fitting
ramp-and-step model is superimposed. Note that, as per Table~1,
varying numbers of PCUs are on for each observation. The ingresses
have been aligned for presentation purposes, and in each case 860~sec of
data have been excluded between the ingress and egress plots.
\label{f-xray}}
\end{figure}

The X-ray observations were scheduled to occur centered on the
expected times of eclipse, as extrapolated from the ephemeris of
Cominsky \& Wood (1989). As intended, one complete eclipse was
observed per observation. We have determined the duration, mid-point
and transition times for each eclipse by modeling each ingress and
egress transition with a ``step and ramp'' model, consistent with
the methodology
adopted in studies of eclipses from the similar transient LMXB
X0748$-$676 (e.g. Parmar et al.\ 1986, Corbet et al.\ 1994). The model
assumes a linear transition into and out of eclipse, and
has four free parameters per transition:
the start and end time, and the count
rates before and after the transition.  From these we derive the
ingress and egress durations, $\Delta T_{ing}$ and $\Delta T_{egr}$,
the eclipse duration $\Delta T_{ecl}$ (measured from the end of
ingress to the beginning of egress), and the eclipse mid-points
(midway between the end of ingress and the start of egress).  Table~2
contains the measured values of these quantities for each eclipse. We
find a spread of ingress/egress times of 6--13~sec,
with mean values for $\Delta T_{ing}$ and $\Delta T_{egr}$ of
9.1$\pm$3.0~sec and 9.5$\pm$3.3~sec respectively, and a mean eclipse
duration of 901.9$\pm$0.8~sec. The X-ray eclipse transitions of
X1658$-$298 together with the model fits are shown in Figure~\ref{f-xray}.

\begin{figure}[htb]
\plotone{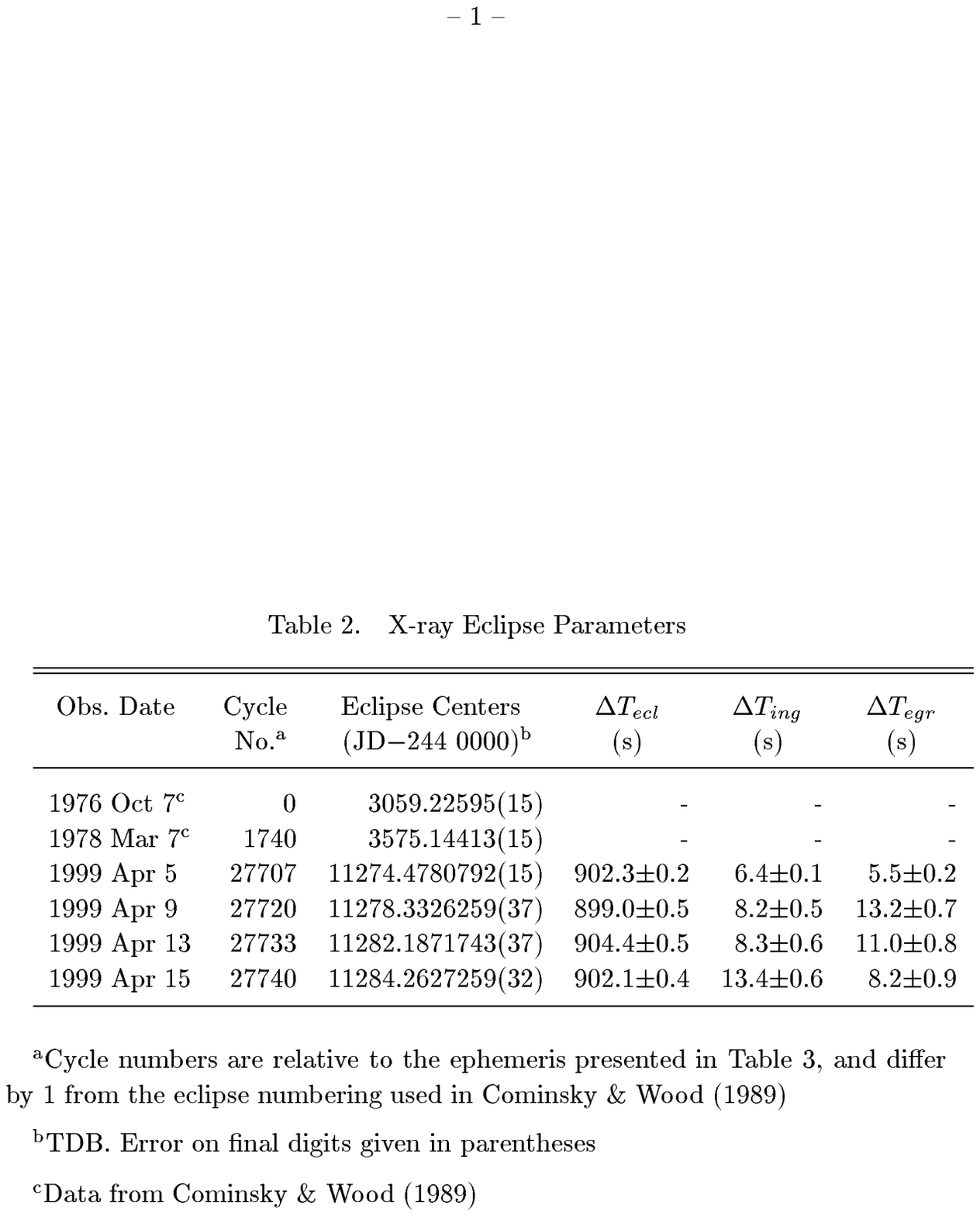}
\end{figure}

These four eclipse centers occur an average of 407.4~sec earlier than
predicted by the ephemeris of Cominsky \& Wood (1989).  We have
combined our eclipse timings with the eclipse centers (corrected to
TDB) from the HEAO A-1 and SAS 3 observations of Cominsky \& Wood
(1984, 1989), to produce the updated ephemeris presented in Table~3.
A parabolic ephemeris is required to obtain a good fit to the eclipse
timings; the $\dot{P}$/$P$ term implies that the orbital period of the
system is decreasing on a timescale of 10$^7$ yr.  This ephemeris was
then used to phase our optical data, in the following sections.

\begin{figure}[htb]
\plotone{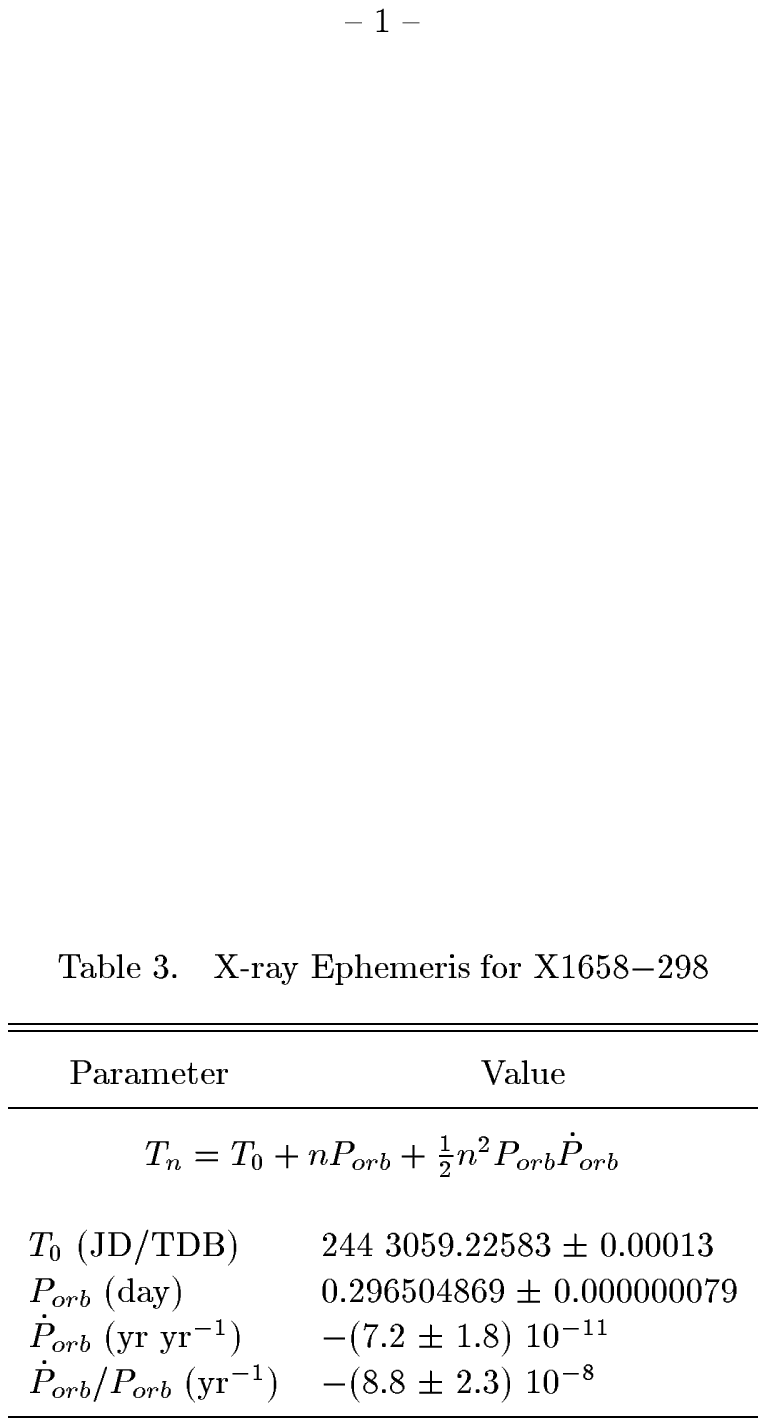}
\end{figure}

We have also performed a spectral analysis of the RXTE PCA data.  From
each dataset, we extracted a spectrum of the persistent (non-eclipse,
non-dip) emission, and an in-eclipse spectrum. The spectra of the
persistent emission each contain $\sim$1300~sec of data, and the
eclipse spectra $\sim$880--896~sec. In each case the persistent
spectrum can be well fit using a power law plus high energy cutoff
model, with power law index $\alpha$=2.1$\pm$0.1, cutoff 8.6$\pm$0.6
keV, and a hydrogen column density $N_H$ of
(5.0$\pm$0.6)$\times$10$^{22}$ cm$^{-2}$. A Comptonized Sunyaev and
Titarchuk model also provides reasonable fits to the data, with
$kT$=3.9$\pm$0.2 keV, $\tau$=7.1$\pm$0.3,
$N_H$=(6.1$\pm$0.5)$\times$10$^{22}$ cm$^{-2}$. The reduced $\chi^2$
values for both models are acceptable, in the range 1.0--1.2. Two
component models (such as a powerlaw plus blackbody) will also fit the
data, although an F-test does not justify the inclusion of the second
component. The mean persistent 2--20 keV flux of the source throughout the
observations is 1.05$\times$10$^{-9}$ erg~cm$^{-2}$~s$^{-1}$.
The eclipse spectra can be consistently fit with a simple, steeper
power law with $\alpha$=3.5$\pm$0.4 and
$N_H$=(15$\pm$5)$\times$10$^{22}$ cm$^{-2}$. 
Over the 2--20 keV range, the
eclipse flux level is measured to be 1.9$\pm$0.7\% of the persistent 
emission.

The durations we measure for the X-ray eclipse transitions are shorter than
those determined from the previous activity cycle (mean $\Delta
T_{ing}$=41$\pm$13~sec, mean $\Delta T_{egr}$=19$\pm$13~sec; Cominsky \& Wood
1989). However, a broad spread of values for the eclipse transitions
from a given source may be common; the similar source X0748$-$676
shows transition times from 1.5--40~sec (Parmar et al.\ 1991). Transition
times are defined by the atmospheric scale height of the companion,
which can be affected by flaring activity or the presence of an
X-ray-induced evaporative wind or corona; a more detailed discussion
of of such effects in X1658$-$298 will be worthwhile once a larger
sample of eclipses is obtained.

Period changes have been previously detected in six other LMXBs, and may
provide valuable clues about the progression of binary evolution.  For
conservative mass transfer, the loss of angular momentum leads to an
expected timescale for evolution of the orbital period
($\tau$=$P_{orb}/\dot P_{orb}$) of 10$^{8-10}$~yr. However, the
timescales measured to date have been considerably shorter than this.
The period of X1822$-$371 and X2127$+$119 are increasing on timescales of 
$\tau$=2.9$\times$10$^6$~yr and $\tau$=1.1$\times$10$^6$~yr,
respectively (Hellier et al.\ 1990; Homer \& Charles 1998), while
X1820$-$303 and Her~X--1 show decreasing orbital periodicities with
$\tau$=1.9$\times$10$^7$~yr (van der Klis et al.\ 1993 and references within)
and $\tau$=7.6$\times$10$^7$~yr (Deeter et al.\ 1991).
Cyg~X--3 (possibly not an LMXB) shows an increasing orbital period,
with $\tau$=7.3$\times$10$^{6}$~yr, with a possible second period
derivative (van der Klis \& Bonnet-Bidaud 1989; Kitamoto et
al.\ 1992). Most complex of all, X0748$-$676 shows a period change
behavior initially seen to decrease (Parmar et al.\ 1991) but later
impossible to reconcile with a simple constant period derivative. A
sinusoidally-varying orbital period (Asai et al.\ 1992) provided an
acceptable fit until the RXTE era, when an unusually large excursion
from this pattern was detected that defies straightforward
parameterization (Hertz, Wood, \& Cominsky 1997).  The variation
observed in X1658$-$298 is of a similar magnitude to these cases,
despite the fact that (presumably) mass transfer was not occurring
during the interval 1978--1999. This may pose a difficulty in
explaining the change using models based on angular momentum coupling,
irradiation of the secondary, or magnetic cycling (e.g. Parmar et al.\ 1991;
Richman, Applegate, \& Patterson 1994; Hertz et al.\ 1997).

\subsection{Optical}

\begin{figure}[ht]
\plotone{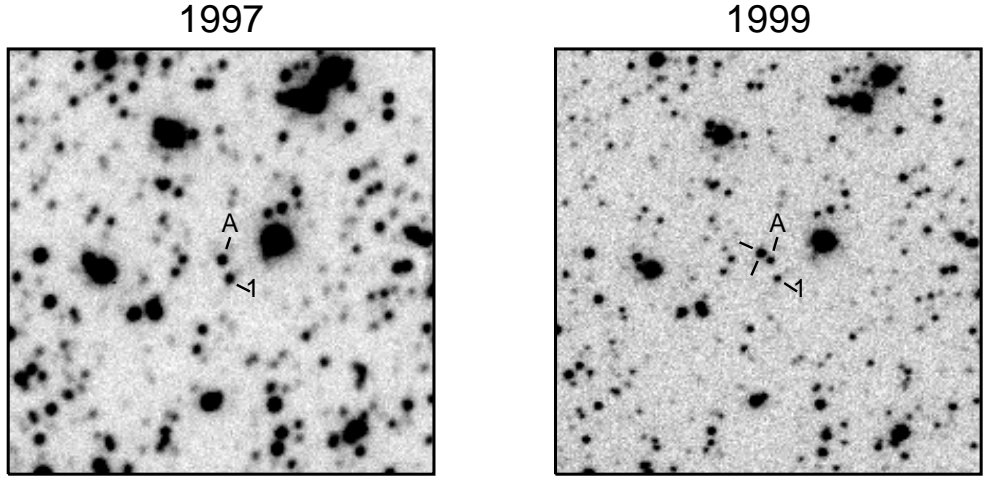}
\caption[]{
50\arcsec $\times$50\arcsec\ $I$ band exposures of the X1658$-$298 field
obtained with the CTIO 1.5m telescope while the source was in
quiescence (left, 1997 May 5) and during outburst (right, 1999 May 2).
Star ``A'' is the star assumed previously to be the counterpart
(Wachter \& Smale 1998). The actual counterpart (flagged) is
0.8\arcsec\ east and 1.0\arcsec\ north of A.
 \label{f-fc}}
\end{figure}

Apart from the discovery observations during the 1978 outburst and
limited follow-up during the subsequent decay, optical
data of V2134~Oph are sparse.
The few more recent spectroscopic observations of V2134~Oph during the 
extended X-ray off--state
(Cowley, Hutchings \& Crampton 1988; Shahbaz et al.\ 1996; Navarro 1996)
all imply a substantially brighter counterpart than the 
$V>23$ limit discussed in Cominsky et al.\ (1983).
We conducted the first photometric study of V2134~Oph in 1997 April/May 
and found the source only $\sim$~1~mag fainter than the 
brightness reported when the system is X-ray active (Wachter \& Smale 1998).
Our lightcurve surprisingly also did not show any modulation across the 
binary cycle. 
X-ray transients in quiescence generally display photometric variability 
due to ellipsoidal variations (see van Paradijs \& McClintock 1995 for 
extensive references). 
A comparison 
between our quiescent and outburst frames from 1997 and 1999 (Figure~\ref{f-fc})
resolves this puzzling behavior: the star we observed (``A'') is  
in fact an unrelated close companion to the actual optical counterpart. 
The true position of
V2134~Oph was measured from a large number of frames with excellent seeing 
conditions (0.7-0.8\arcsec) to be 0.8\arcsec\ east and 1.0\arcsec\ north of the 
star marked ``A''. The presence of this unrelated companion cannot be 
discerned from the original 
finding chart of Doxsey et al.\ (1979). 
There is no evidence for intermittent brightening of the source in X-rays 
during the quiescent years, and consequently it is probable that 
the counterpart remained at the faint $V>23$ magnitudes 
reported by Cominsky et al.\ (1983). This magnitude is 
inaccessible for spectroscopic studies with the telescopes used in 
the observations referenced above and it is likely that none of these
observations were of the true quiescent counterpart. 

We subsequently reanalyzed our 1997 data in order to determine quiescent
magnitudes of the true counterpart which is only very faintly 
visible in our pre-outburst data. After coadding the six $I$ frames (600 sec 
each) with the 
best seeing conditions, we obtain $I=22.1 \pm 0.3$ for 
V2134~Oph in quiescence. Unfortunately, the quiescent counterpart is  
too faint for accurate photometry in our 1997 $V$ and $R$ band data.
Filippenko et al.\ (1999) measured $R=23.6\pm 0.4$ for V2134~Oph in quiescence.
Combining these two measurements and 
assuming a reddening of $E_{B-V}$=0.3 (van Paradijs 1995) results 
in $(R-I)_0= 1.2$ which corresponds to a
spectral type of M2. A star of this spectral type would not fill its Roche 
lobe in a 7.1~hour LMXB. As discussed in Wachter \& Smale (1998), 
empirical period--mass relations for mass transfer systems imply a K0 star 
instead. 
Note, however, that the $R$ magnitudes of the comparison stars A and E 
(A and 1 in 
our nomenclature) in Filippenko et al.\ (1999) are systematically 0.6-0.7 
mag fainter than those given
in Wachter \& Smale (1998). If the $R$ magnitude of V2134~Oph is similarly too 
faint, the resulting $(R-I)_0$ is consistent 
with the required early K spectral type within the errors. A K0 companion would have 
$V=23.6$, in accordance with the observed $V>23$ limit.   

\begin{figure}[htbp]
\plotfiddle{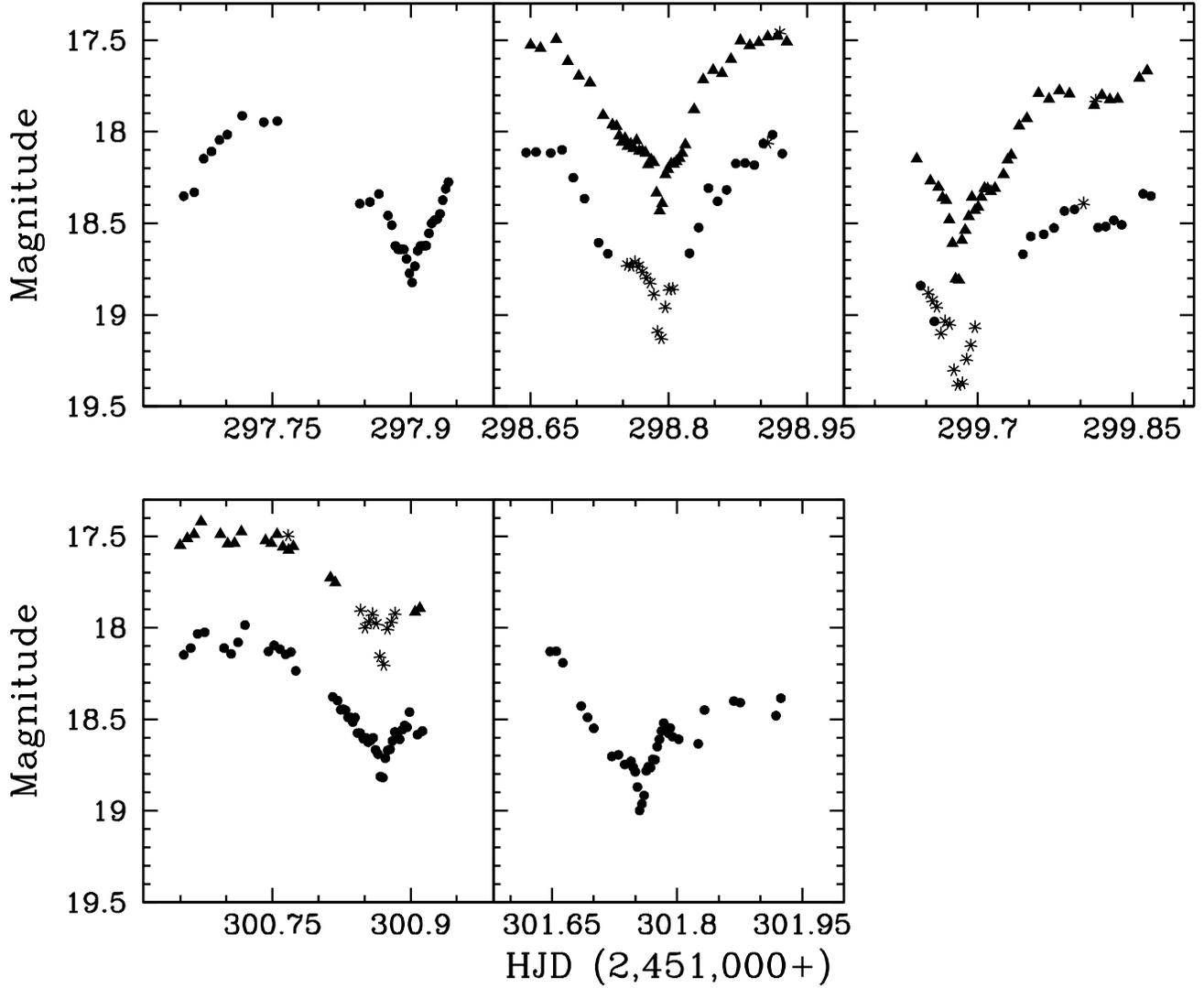}{15cm}{-90.}{90.}{90.}{-350}{500}
\caption[]{
$V$ and $I$ band lightcurves of V2134~Oph, the optical counterpart of
X1658$-$298,
obtained in 1999 May, shortly after renewed X-ray activity was detected
from the source.
Data obtained with the CTIO 1.5m telescope
are indicated with filled circles ($V$) and triangles ($I$), data obtained
with the YALO 1m telescope
are indicated with stars.
 \label{f-lc}}
\end{figure}
 
Our 1999 $V$ and $I$ band lightcurves of V2134~Oph during outburst are
shown in Figure~\ref{f-lc}. Data obtained with the CTIO 1.5m telescope
are indicated with filled circles for $V$ and triangles for $I$, YALO 
data with stars. The observations span 
almost a full orbital cycle on each night. Superposed on a gradual brightness
variation with $\sim 0.7-0.8$~mag amplitude, a distinct, narrow eclipse 
feature of about 0.2~mag is visible on each night.  
Strong nightly variations in 
the shape of the lightcurve are also evident. 
For the nights with simultaneous $V$ and $I$ band coverage we
rebinned the data to the average time sampling interval using 
linear interpolation and calculated the $(V-I)$ color index.  
There is no evidence for any 
$(V-I)$ color variation across the orbit or for a change in color from 
night to night. 
We obtain an average of
$(V-I)=0.645 \pm 0.054$ from the combined color data of the three nights.

\section{Discussion}

\begin{figure}[htbp]
\epsscale{0.8}
\plotone{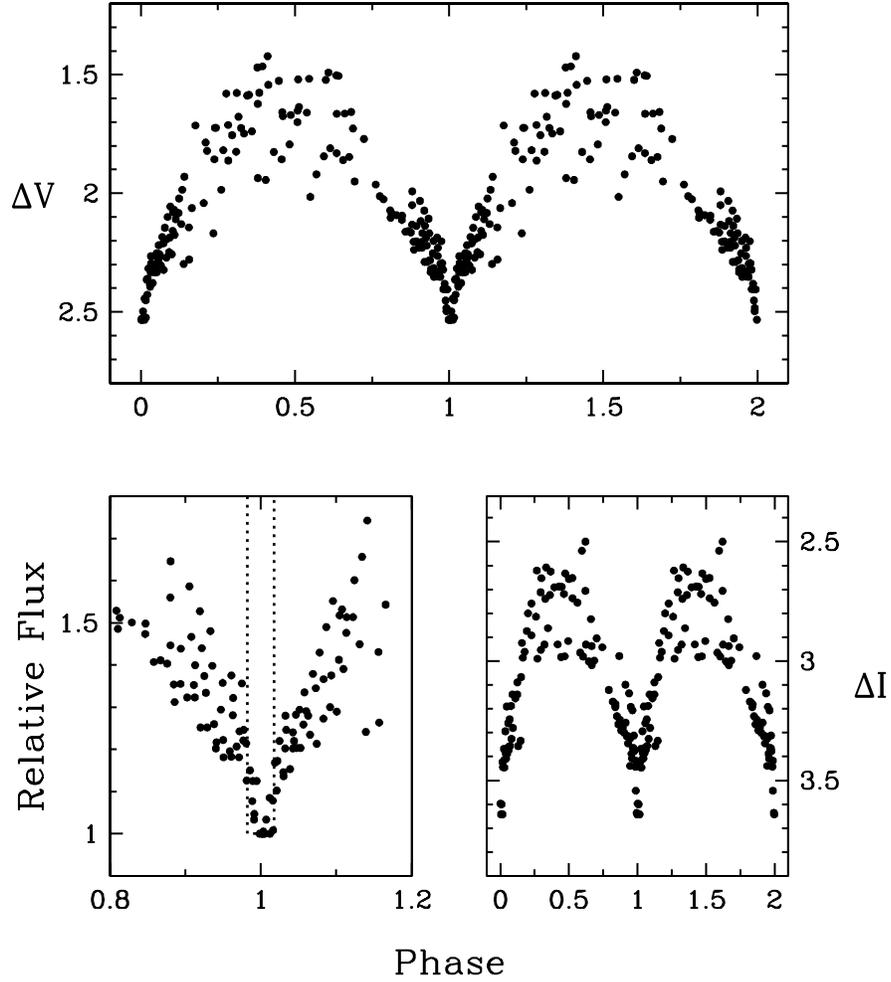}
\caption[]{
Our $V$ and $I$ band data of V2134~Oph folded according to our updated
X-ray ephemeris. A close-up of the $V$ band eclipse center is shown
on the bottom left hand side of the plot together with the fit to the
1999 Apr 5 X-ray eclipse (dotted line).
In the usual convention, the time of the X-ray
mid-eclipse is defined as phase 0. The deepest point of the eclipse on
each night
was chosen as a reference point for the brightness of the system and the
data (vertically) shifted accordingly.
\label{f-flc}}
\end{figure}

Figure~\ref{f-flc} shows the outburst data folded according to our 
updated X-ray ephemeris 
(Table~3). Following the usual convention, the time of the X-ray 
mid-eclipse is defined as phase 0. The deepest point of the 
optical lightcurves on each night 
was chosen as a reference point for the brightness of the system and the 
data (vertically) shifted accordingly. 
X-ray dips are observed for X1658$-$298 
between the phases of 0.6--0.8. No analogous stable optical feature 
is evident in the folded lightcurves, however, our data sampling in that
phase interval is fairly sparse. The folded $V$ band lightcurve clearly shows 
a distinct central drop in brightness within $\sim 0.2$~mag of the 
faintest observed magnitude which is also characterized by reduced scatter
compared to other phases of the lightcurve. The presence of such a narrow 
central component is evident in the individual lightcurves of each 
separate night as well. 
We determined the optical eclipse center to occur at phase $0.004 \pm 0.003$
by selecting the folded $V$ band data within 0.2~mag of the faintest 
magnitude and calculating the time on either side of which the area 
within the eclipse profile was equal.
The average data sampling interval in this part of
the folded $V$ band lightcurve is about 1 minute. A close-up of the central 
region is shown on the bottom left hand side of Figure~\ref{f-flc} together 
with the fit to the 1999 Apr 5 X-ray eclipse (dotted line). 
Our data indicate that
there is no significant offset between the time of mid-eclipse in the 
X-ray and optical and that the  
narrow component of the optical eclipse is of the same duration as the 
X-ray eclipse (we measure a
FWHM of $14 \pm 2$ minutes for this optical feature).
This implies a distinct optical emission region 
associated with the X-ray emitting area. 

The only other LMXBs known to exhibit X-ray eclipses are 
X0748$-$676, X2129+470, X1822$-371$, Her X$-$1 and X0921$-$630. 
For systems with inclinations $75^\circ \lesssim i \lesssim 80^\circ$
(X1658$-$298, X0748$-$676, Her X$-$1), 
both dips and total eclipses are observed in X-rays. In
higher inclination systems ($i \gtrsim 80^\circ$; X1822$-371$, X2129+470), 
only partial X-ray eclipses are seen; the accretion disk is thought to block 
the direct line-of-sight to the central X-ray source and the observed X-ray
flux is due to scattering in an extended accretion disk corona (ADC).
The optical/UV eclipse in the ADC source X1822$-$371, one of the most
extensively studied systems, is found to be much 
broader than the X-ray eclipse (Hellier \& Mason, 1989; Puchnarewicz, 
Mason, \& Cordova 1995) indicating an accretion disk radius of about 
twice the ADC radius. The optical lightcurve of X1822$-$371 varies very little
from night to night and even over a timespan of years. Modeling
shows that several emission components such as the X-ray heated face of
the mass donor and the accretion disk rim contribute to produce the
overall morphology of the optical lightcurve (Mason \& Cordova 1982). 
It is therefore difficult
to determine the time of ingress and egress of the
optical eclipse for a given system solely from the shape of the lightcurve.

In contrast to X1822$-$371, our X1658$-$298 data clearly display a
narrow optical feature of the same duration as the X-ray eclipse.
However, the data do not reveal whether this feature merely represents
the central core of a wider optical eclipse. Due to the highly variable
shape of the lightcurve outside this narrow component, we cannot ascertain 
the presence or absence of wider ingress/egress signatures. 
The standard
model calls for successively longer eclipse durations when moving from 
observations in X-rays to longer wavelengths to account for the eclipse 
of the cooler outer regions of an extended accretion disk (which would not 
be visible in X-rays).
Our data clearly indicate an accretion disk structure characterized
by enhanced optical emission coincident with the central X-ray emitting area. 
We would consequently predict equivalent optical structures in all systems
in which the X-ray source is believed to be viewed directly. 
An optical feature on the timescale of the X-ray eclipse has been observed
in Her~X$-$1 (Kippenhahn, Schmidt, \& Thomas 1980). For X0748$-$676,
an inspection of the individual optical lightcurves displayed in
Crampton et al.\ (1986) and van Paradijs, van der Klis, \& Pedersen (1988) 
also reveals a narrow central eclipse component very similar to that of 
our X1658$-$298 data. However, in both cases
the authors conclude that the optical eclipse is twice as wide as the X-ray 
eclipse, based on which part of the lightcurve looks like ``clearly an 
eclipse'' and/or consideration of an average lightcurve which does not 
exhibit any central structure. While it is difficult to tell with 
certainty from the 
published figures, it appears likely that a reexamination of 
these X0748$-$676
data would also show this narrow optical component to have the same
duration as the X-ray eclipse.

\acknowledgments

We thank Alistair Walker for assigning us director's 
discretionary time for the optical observations of this project.
This research has made use of the Simbad
database, operated at CDS, Strasbourg, France and 
of results provided by
the ASM/RXTE team at MIT and NASA/GSFC.
C. Bailyn is supported by NSF AST 97-30774.

\end{document}